# Computational Analysis of Interface-Driven Spin-Orbit Coupling in Molecular Adsorbates on Transition Metal Dichalcogenides

Zihao Wang,[1,2] Wan-Lu Li,[2,3,†] and Shaowei Li[1,2,†]

[1]*Department of Chemistry and Biochemistry, University of California, San Diego, La Jolla, CA 92093-0309, USA*
[2]*Program in Materials Science and Engineering, University of California, San Diego, La Jolla, CA 92093-0418, USA*
[3]*Aiiso Yufeng Li Family Department of Chemical and Nano Engineering, University of California, San Diego, La Jolla, CA 92093-0448, USA*

Spin-orbit coupling (SOC) lifts molecular orbital degeneracy, enabling bi-level electronic platforms suitable for next-generation digital devices. However, common light-atom molecular feedstocks exhibit weak SOC due to the absence of heavy elements. To enhance SOC without synthesizing new materials, we leverage interfacial interactions between molecules and transition-element-based solid-state materials. This computational study investigates SOC splitting in metal-phthalocyanine adsorbed on transition metal dichalcogenides (TMDs) using density functional theory (DFT). The enhanced SOC splitting is attributed to strong orbital hybridization at the molecule-substrate interface. Specifically, Zn-phthalocyanine (ZnPC) on monolayer $MoS_2$ achieves a notable SOC splitting of ~8 meV. Furthermore, when ZnPC forms self-assembled chains on $MoS_2$, the splitting increases to ~20 meV, driven by the formation of hybrid bands modulated by molecular periodicity. These findings highlight the role of interfacial and intermolecular interactions in inducing and enhancing SOC in surface-adsorbed molecules, providing a new strategy for molecular spintronic materials without complex synthetic efforts.

*Introduction*—Molecular spintronics focuses on using electron spin in molecular orbitals for information processing and storage [1-3]. Enhancing spin-orbit coupling (SOC) is key for manipulating spin states via external electric fields [4-8], influencing spin state energy degeneracy [9,10] and relaxation times [11-15], which affects the coherence and transport of spin information. SOC enables efficient spin control via electric dipole spin resonance [16-19] and is tunable by modifying molecular structures [13], providing freedom of reconfiguration in spintronic devices. However, common molecular materials, often only containing light atoms, have weak intrinsic SOC, prompting extensive research efforts to incorporate heavier elements or form transition metal complexes. These compounds, though effective, are often either synthetically challenging or chemically unstable, limiting their broader applications [20-22].

Alternatively, interfacial engineering can enhance SOC in molecules without complex synthesis by leveraging existing solid-state materials with intrinsic strong SOC [1]. By depositing molecules onto substrates like transition metal dichalcogenides (TMDs) [23], interactions with heavy transition metals in substrates promise to induce SOC in the molecular orbitals. TMDs, particularly monolayers, exhibit strong SOC due to a lack of spatial inversion symmetry, leading to the spin-polarized band splitting [24-26]. This unique property has advanced spin-valley electronics [27-29], the quantum spin Hall effect [30-32], and spin-optoelectronics [33-37]. TMDs, especially in van der Waals (vdW) heterostructures, are ideal for inducing SOC in adjacent layers [11,38-42], as demonstrated in $WS_2$-graphene heterojunctions [12,43-46]. Similarly, experimental techniques enable the deposition of molecular adsorbates on TMDs, which is expected to induce SOC in the adsorbates [47].

In this study, we computationally explore the interfacial interactions between molecules and various TMD substrates to assess interface-induced SOC in molecular adsorbates, specifically focusing on heterojunctions exhibiting SOC-induced energy splitting at zero magnetic fields. Due to the relatively weak SOC Hamiltonian in molecules compared to the Coulomb potential, zero-field SOC splitting in molecules is rare in nature [13,14,48]. We selected metal phthalocyanine (metal-PC) as a model molecule for its known planar structure that promotes strong substrate contact together with its ability to self-assemble on TMDs [49,50]. Using density functional theory (DFT) simulations, we systematically screened metal-PC/TMD heterojunctions, employing spin-polarized density of states (SP-DOS) and wavefunction analyses to evaluate SOC splitting energy. Our results show that the ZnPC-$MoS_2$ heterojunction exhibits the most pronounced SOC-induced orbital splitting, without contributions from Zn d-electrons. Additionally, we find that the self-assembly of ZnPC on $MoS_2$ results in molecular arrays that further enlarge the energy splitting that is expected to be readily measurable [51]. This study provides theoretical guidance for the further exploration of heterogeneous spintronic materials.

*SOC Hamiltonian in molecules*—Fundamentally, the SOC Hamiltonian is expressed with the Dirac equation as:

$$H_{SOC} = \frac{1}{2m_e^2 c^2} (\nabla V \times \boldsymbol{p}) \cdot \boldsymbol{S}. \qquad (1)$$

For an isolated molecule, the total SOC ($H_{SOC}$) is obtained as the sum of the SOC Hamiltonians of individual atoms. Using central potential approximation, the $H_{SOC}$ is:

$$H_{SOC} = \frac{1}{2m_e^2 c^2} \sum_n \left(\frac{1}{r}\frac{dV}{dr}\right)_n \boldsymbol{L}_n \cdot \boldsymbol{S} \qquad (2)$$

---

[†]Contact author: wal019@ucsd.edu
[†]Contact author: shaoweili@ucsd.edu

where $\Sigma_n$ represents the summation over all atoms. However, if molecules form a periodic lattice on top of a solid-state material surface, localized molecular orbitals hybridize with the delocalized substrate bands. In this case, we need to decompose the electron potential in a molecule into two components:

$$V = V_{\text{Central field}} + V_{\text{nonCentral field}} \quad (3)$$

where the first term remains a localized central potential field around each atomic nucleus within the individual molecule, while the second term represents a translational invariant defined by the periodicity of the molecular lattice. Therefore, the SOC Hamiltonian can now be expressed as:

$$H_{\text{SOC}} = \frac{1}{2m_e^2 c^2} \sum_n \left(\frac{1}{r} \frac{dV_{\text{Central field}}}{dr}\right)_n \mathbf{L}_n \cdot \mathbf{S}$$
$$+ \frac{1}{2m_e^2 c^2} (\nabla V_{\text{nonCentral field}} \times \hbar \mathbf{k}) \cdot \mathbf{S} \quad (4)$$

where we can define a nonzero wave vector $\mathbf{k}$ within the first Brillouin zone of the molecular lattice.

To enhance the molecular SOC, we seek to maximize each term in $H_{\text{SOC}}$ (Eq. 4). The first term is linked to the product of an electron's orbital ($\mathbf{L}$) and spin ($\mathbf{S}$) angular momentum. In organic molecules composed of main-group elements, electronic orbitals are predominantly s and p types, resulting in minimal $\mathbf{L}$ and weak SOC. In contrast, TMDs with heavy transition metals like molybdenum and tungsten have strong SOC from d and f orbitals. Interfacial hybridization, when molecules are deposited on TMDs, can promote interaction between the transition metal and molecular electrons, increasing SOC energy. Additionally, interfacial hybridization delocalizes molecular orbitals, enabling interactions between nearby molecules through the substrate band. This makes $V_{\text{nonCentral field}}$ in the 2nd term of $H_{\text{SOC}}$ in Eq. 4 significant, suggesting that forming a molecular lattice can further elevate the spin degeneracy of the molecular orbitals to establish the bi-level quantum system.

*Interfacial proximity-induced SOC splitting in ZnPC*— Following these hypotheses, we selected ZnPC and MgPC as test molecules for the DFT study due to their closed-shell electronic configurations of the metal centers. All DFT calculations are implemented in the Vienna *ab initio* simulation package (VASP) [52-54]. The projector augmented wave (PAW) method [55] and the generalized gradient approximation (GGA) of the Perdew-Burke-Ernzerhof (PBE) [56,57] exchange correlation functional are adopted. Details of the computational methods are introduced in the Supplemental Material (SM) part 1 [52-59]. In both gas-phase ZnPC and MgPC, the highest occupied (HOMO) and lowest unoccupied (LUMO) molecular orbitals do not involve contributions from the metal center, thereby excluding SOC effects from the metal itself. Notably, without considering SOC, the LUMO consists of two degenerate orbitals, noted as $|x\rangle$ and $|y\rangle$ in the SM Fig. S1. Notably, they are no longer a good basis set after including the SOC Hamiltonian. SOC induces mixing between $|x\rangle$ and $|y\rangle$, as well as between spin-up and down states, resulting in negligible intrinsic energy splitting below 1 meV (SM part 2).

To explore interfacial interactions between molecules and heavy-element substrates, we use Mo(100), a simple monoatomic crystal with a rich d-band near the Fermi level, as the testbed substrate. We simulate ZnPC orbital splitting at varying molecule-substrate distances to assess interfacial SOC strength. A $7 \times 7$ bilayer Mo supercell models the substrate, with self-consistent-field (SCF) calculations performed on a $\Gamma$-centered $1 \times 1 \times 1$ Monkhorst-Pack k-point grid. The Mo surface remains paramagnetic to ensure zero total magnetic moment. As shown in Fig. 1a and 1b, ZnPC adopts planar adsorption geometry at a fourfold symmetric site [60,61], minimizing symmetry-breaking effects on molecular orbital energy level splitting. At 0.6 Å away from the relaxed adsorption geometry, the LUMO orbital of ZnPC exhibits a SOC-like splitting of 8.8 meV (Fig. 1c-d). The SP-DOS retains a symmetric distribution around zero, consistent with the preserved time-reversal symmetry (TRS). This observation indicates that the interfacial interaction can generate detectable molecular SOC splitting. However, at the stable configuration, the ZnPC-Mo(100) distance is much smaller, leading to strong hybridization that diminishes molecular orbital characteristics. While SOC-induced splitting increases as ZnPC nears the surface (black line, Fig. 1e), significant charge transfer (red line, Fig. 1e) disrupts molecular orbitals and shortens excited state lifetimes.

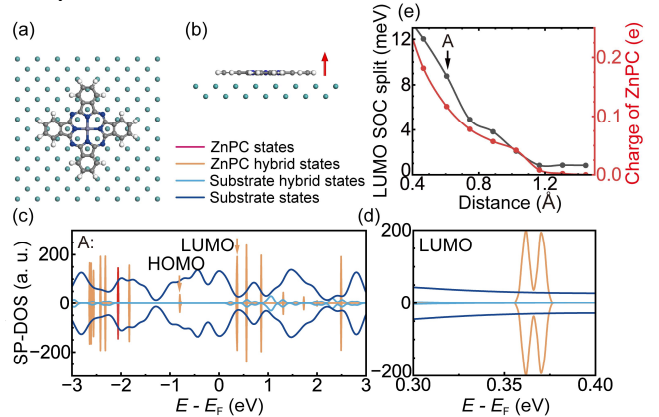

FIG. 1. (a-b) The top (a) and side views (b) of the relaxed ZnPC adsorption configuration on Mo(100). (c) SP-DOS of ZnPC adsorbed on Mo(100) with the molecule positioned 0.6 Å away from the relaxed adsorption structure, corresponding to the distance labeled as 'A' in (e). Different colors represent electronic states with varying degrees of hybridization: monomeric and hybrid ZnPC orbitals are shown in red and orange, respectively, while monomeric and hybrid substrate states appear in dark and light blue. (d) Simulated energy profile of the ZnPC LUMO. (e) SOC-induced LUMO splitting (black) and interfacial charge transfer (red) at different molecule-substrate separations. The distance zero corresponds to the relaxed adsorption geometry.

*TMD-induced metal-PC SOC splitting*—Given that the hybridization between the metal substrate and the molecule is too strong, smearing the molecular orbital characters, molecule-TMD heterojunctions emerge as a better system to mitigate this issue. TMDs weakly interact with the adsorbed molecule while also containing transition metals that can be leveraged to enhance SOC in molecular adsorbates. We simulate the interaction of ZnPC and MgPC on monolayer TMDs, including $MoS_2$, $MoSe_2$, $WS_2$, and $WSe_2$, to determine the relative alignment between the molecular HOMO-LUMO gap and the TMD band gap.

Previous reports have shown that ZnPC undergoes planar vdW adsorption on the TMD surface, exhibiting negligible energy differences among various adsorption configurations [49,50]. Since the adsorption site has minimal influence on the interaction between planar metal-PC molecules and TMD states, a $6 \times 4\sqrt{3}$ supercell of monolayer TMD is chosen, with the molecule's metal center positioned at the hollow site, as shown in Fig. 3a. A Γ-centered Monkhorst-Pack k-point grid of $1 \times 1 \times 1$ was used for SCF calculations. This setup reduces computational effort in identifying the global minimum energy configuration while preserving the integrity of the molecular and TMD energy levels.

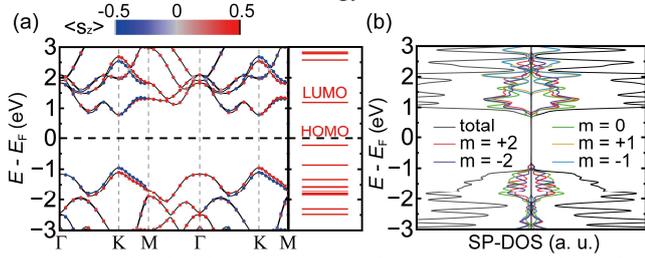

FIG. 2. (a) Simulated band structure of $MoS_2$ monolayer including SOC (left panel) and the energy levels of gas-phase ZnPC (right panel). The color denotes the spin expectation along the z-axis. (b) Orbital-projected DOS of monolayer $MoS_2$, considering the d-orbitals of Mo atoms with different magnetic quantum numbers $m$.

Among all the TMD substrates we simulated, the ZnPC-$MoS_2$ heterostructure exhibits the most significant interface-induced LUMO splitting, attributed to the favorable energy alignment between the LUMO and the conduction band (CB) of $MoS_2$. The energy alignment between the monolayer $MoS_2$ band structure and the isolated ZnPC energy levels is shown in Fig. 2a. The simulated result exhibits a ~147 meV SOC splitting near the K point of the valance band (VB) in $MoS_2$, closely matching the experimental value of 150 meV [62]. Similarly, the lowest CB also undergoes SOC splitting of ~60 meV along the Γ-K and K-M path. Fig. 2b displays the SP-DOS and Mo d-orbital-projection DOS for monolayer $MoS_2$. The VB maximum (VBM) is primarily composed of $m = \pm 2$ orbitals, while the CB minimum (CBM) is dominated by $m = 0$ orbital, accounting for the weak SOC at the CB edge. However, as energy increases, the DOS of $m = \pm 2$ orbitals rise rapidly, leading to a significant SOC splitting in CB.

The SP-DOS for the ZnPC-$MoS_2$ heterostructure is presented in Fig. 3b and 3c. The LUMO exhibits an interface-induced SOC splitting of 8.4 meV while preserving TRS. This energy split is significantly larger than the intrinsic SOC effect in isolated ZnPC, suggesting that interaction with the TMD significantly enhances the SOC in adsorbed molecules. The k-space distribution of peaks $a$ and $b$ from Fig. 3c reveals that the TMD states interacting with the LUMO are primarily located near the K-M and Γ-K paths in the CB (Fig. S6), where pronounced SOC splitting is observed (Fig. 2a). The calculated g-factors of peaks $a$ and $b$ are 1.75 and 1.81, respectively, both lower than the free-electron g-factor ($g_e$ = 2.00). This reduction results from the opposite orientations of their orbital and spin magnetic moments. The orbital magnetic moments of peaks a and b primarily originate from Mo d-orbitals. Their orbital moments are nearly identical, stemming from their small energy splitting, as both states interact with the same $MoS_2$ band. The SOC-induced energy splitting between peaks $a$ and $b$ is mainly attributed to molecular spin reorientation (SM part 7 and Fig. S9).

Notably, the band gap calculated using the GGA functional is lower than the experimental value [63,64]. This discrepancy arises from the known tendency of the GGA functional to underestimate electronic band gaps. However, this offset does not compromise the accuracy of the simulated SOC, as the distortion in GGA exchange interactions primarily occurs near the atomic core, whereas the molecule-TMD interaction is primarily long-range. To confirm this, we have calculated the electronic structure of this heterostructure using Heyd-

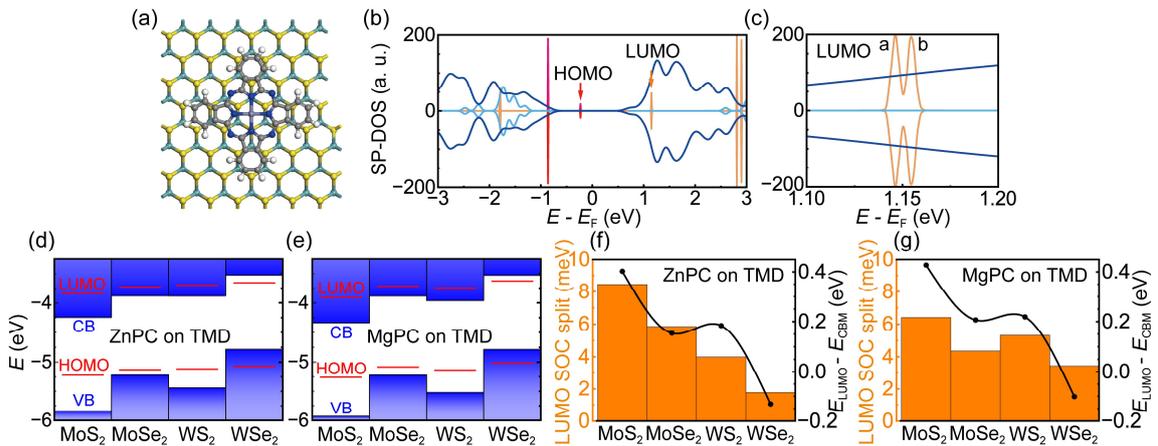

FIG. 3. (a) Schematic configuration of ZnPC adsorbed on a monolayer $MoS_2$. (b) SP-DOS of ZnPC adsorbed on a $MoS_2$ monolayer. (c) Magnified view of the DOS, highlighting the SOC-splitting ZnPC LUMO on $MoS_2$. (d-e) Energy level alignment of the HOMO/LUMO in different metal-PC/TMD heterojunction, with the zero energy set to the vacuum level. (f-g) SOC splitting energy (orange) of the ZnPC (f) and MgPC (g) LUMO, along with the energy difference between the LUMO and the TMD CBM (black) for different TMDs.

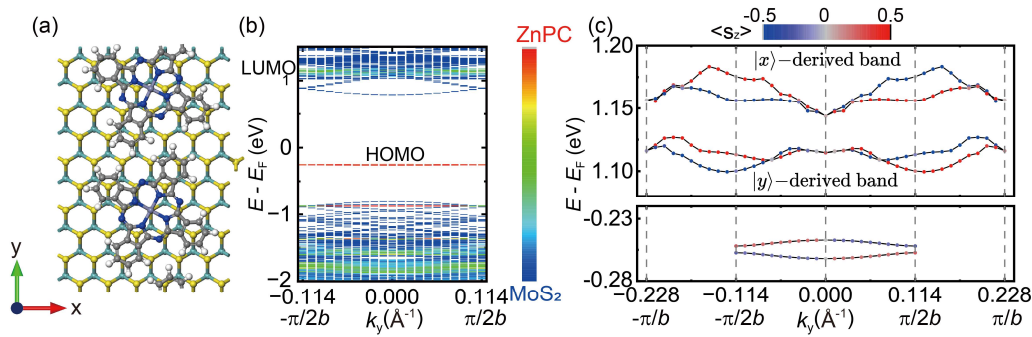

FIG. 4. (a) Calculated structure of the 1D self-assembled ZnPC chain on a MoS$_2$ monolayer. (b) Band structure of the self-assembled ZnPC chain on MoS$_2$. The color of the data points represents the degree of hybridization: red for pure ZnPC states, blue for pure MoS$_2$ states, and intermediate colors for hybridized states. '$b$' denotes the size of one molecule. (c) Magnified band structures of the LUMO-derived band (upper panel) and HOMO-derived band (lower panel), with data point colors indicating the spin expectation along the z-axis.

Scuseria-Ernzerhof (HSE06) hybrid functional [65], which renders a 1.96 eV band gap of MoS$_2$ that closely matches the experimental value. Meanwhile, the LUMO retains the same splitting pattern and magnitude as observed in the GGA calculations (Fig. S7).

Furthermore, the TMD surface lacks the fourfold rotational symmetry of the molecule, which may also contribute to the observed LUMO splitting. To assess this effect, we performed calculations without SOC for the ZnPC-MoS$_2$ heterojunction. The resulting DOS reveals that symmetry breaking induces a splitting of 2.2 meV (Fig. S8), which is significantly smaller than the total splitting after considering SOC. Furthermore, rotational symmetry breaking alone does not cause a g-factor shift. These findings confirm that the splitting observed in Fig. 3c is predominantly driven by the interface-induced SOC.

In addition to ZnPC-MoS$_2$, several other molecule-TMD heterojunctions also exhibit SOC-induced splitting in the LUMO. Generally, the SOC splitting correlates with the DOS overlap between the TMD monolayer CB/VB and the molecular LUMO/HOMO shown in Fig. 3d-e. The magnitudes of LUMO SOC splittings and the energy differences between the LUMO and the CBM are illustrated in Fig. 3f-g. The LUMO splitting shows a positive correlation with the energy difference. This trend arises because both SOC splitting and the density of d orbitals with nonzero magnetic quantum numbers ($m = \pm 2$) in TMDs increase at higher energy (Fig. 2a-b), leading to a stronger LUMO split. The only exception occurs when comparing ZnPC-MoSe$_2$ and ZnPC-WS$_2$, where a larger SOC splitting is observed in ZnPC-MoSe$_2$, while the LUMO is closer to the CBM. A possible explanation for this exception is the placement of the ZnPC LUMO within the DOS valley of the WS$_2$ CB (Fig. S3). The reduced DOS at this location weakens the orbital overlapping, resulting in a smaller induced SOC splitting.

*SOC in 1D array ZnPC on MoS$_2$*—Though ZnPC-MoS$_2$ heterojunction exhibits zero-field splitting due to the interface-induced SOC, it remains relatively weak (in the order of a few meV). As discussed previously, creating an ordered molecular structure may utilize the non-central potential in the SOC Hamiltonian to further create interfacial SOC. Following this idea, we calculated the electronic structure of one-dimensional (1D) self-assembled ZnPC on MoS$_2$. The unit cell employed in our calculations (Fig. 4a) is consistent with both experimental observations [66,67] and theoretical predictions [50]. A $6 \times 5\sqrt{3}$ supercell of monolayer MoS$_2$ was used, containing two molecules per unit cell. The molecular chain direction aligns with the y-axis, with the unit cell defined as $2b$ in this direction. A Γ-centered Monkhorst-Pack k-point grid of $6 \times 4 \times 1$ was used for SCF calculations.

The band structure of the 1D ZnPC chain on MoS$_2$ is shown in Fig. 4b, with the color of the data points indicating contributions from different electronic states. The results reveal weak hybridization between the HOMO and MoS$_2$ orbitals, along with a narrow bandwidth for the HOMO-derived band. In contrast, the LUMO-derived band exhibits a strong hybrid interaction with the MoS$_2$ CB, accompanied by a significantly wider bandwidth. It suggests that direct hopping between adjacent molecules is relatively weak, whereas the substantial overlap between molecular orbitals and the substrate states extends the molecular orbitals. This extension facilitates intermolecular hopping, thereby enhancing the Rashba effect and SOC splitting.

Fig. 4c provides a zoomed-in view of the spin-polarized characteristics of the LUMO- and HOMO-derived bands. The LUMO-derived bands are unfolded due to the presence of two molecules in the unit cell (Fig. 4a), resulting in the duplication of the single-molecule Brillouin zone. Due to the hopping Hamiltonian between neighboring molecules, the degeneracy of the LUMO $|x\rangle$ and $|y\rangle$ is strongly lifted, deriving two distinct bands with an energy separation of 30 meV at $k_y = 0$. Both bands exhibit SOC-induced splitting characteristics independently, analogous to the Rashba effect and the spin-valley locking in TMD band structures. The slight splitting of the HOMO-derived band at $k_y = \pm\pi/2b$ arises from the subtle differences in the chemical environments of the two molecules within the unit cell. Since the HOMO does not hybridize with the TMD states, this small splitting remains visible.

Notably, the LUMO-derived bands exhibit enhanced SOC splitting at $k_y = \pm\pi/2b$, while no splitting is observed at the Γ point ($k_y = 0$) or the Brillouin zone boundary ($k_y = \pm\pi/b$). This behavior arises from the spatial inversion symmetry within the molecular chain. At $k_y = 0$ and $k_y = \pm\pi/b$, the molecular chain forms bonded and anti-bonded states, respectively, preserving the spatial inversion symmetry and reducing SOC effects at these points. In contrast, at $k_y \sim \pm \pi/2b$, the spatial inversion symmetry is maximally broken, leading to a maximum SOC splitting of ~20 meV (SM part 8). For comparison, the HOMO-derived band, weakly coupled to the substrate, remains spin-up and spin-down degeneracy (SOC splitting < 1 meV). It highlights the essential role of molecule-substrate orbital interaction in enabling experimentally observable molecular SOC splitting. The TRS of the system mandates that a state is degenerate with its counterpart possessing reversed momentum and opposite spin. This implies that the hopping current direction of electrons within the molecular chain is locked to their spin-polarization orientation.

*Conclusion*—This study emphasizes the key role of interfacial interactions in enhancing SOC in molecular adsorbates, using metal-PC on TMD substrates as a model system. Our computational screening shows that molecular orbitals interacting with heavy transition metals in the solid-state substrate led to measurable energy splitting. ZnPC-$MoS_2$, in particular, exhibits the strongest SOC-induced splitting in this LUMO orbital, reaching ~8 meV for individual molecules and ~20 meV in self-assembled 1D molecular chain, similar to the spin-valley locking effect observed in monolayer TMDs. This SOC enhancement shifts the g-factor and enables interface-driven spin modulation, providing a theoretical foundation for the design of molecular spintronic devices. These findings present a new strategy for creating molecular spintronic materials by leveraging interface-induced SOC in common molecular feedstocks without complex synthesis.

*Acknowledgments*—This work was primarily supported by the United States National Science Foundation (NSF) under Grant No. 2303936 (to Shaowei Li). This work also used the Expanse supercomputer at the San Diego Supercomputing Center through allocation CSD799 from the Advanced Cyberinfrastructure Coordination Ecosystem: Services & Support (ACCESS) program, which is supported by NSF grants No. 2138259, No. 2138286, No. 2138307, No. 2137603, and No. 2138296.